\documentclass[12pt]{article}%

\usepackage[authoryear]{natbib}

\usepackage{amssymb}
\usepackage{amsfonts}
\usepackage{amsmath}
\usepackage[nohead]{geometry}
\usepackage[singlespacing]{setspace}
\usepackage{indentfirst}
\usepackage{endnotes}
\usepackage{graphicx}%
\usepackage{rotating}
\usepackage{verbatim}
\usepackage{setspace}
\usepackage{multirow}
\usepackage{latexsym}

\usepackage{amsthm}
\usepackage{makeidx}

\newcommand{\cov}{\mathrm{cov}}

\newcommand{\beq}{\begin{eqnarray*}}
\newcommand{\eeq}{\end{eqnarray*}}

\numberwithin{equation}{section}
\theoremstyle{plain}

\newtheorem{cor}{Corollary}[section]
\newtheorem{prop}{Proposition}[section]
\newtheorem{assum}{Assumption}[section]
\theoremstyle{definition}

\makeatletter
\def\@biblabel#1{\hspace*{-\labelsep}}
\makeatother
\begin{document}

\title{A Note on Choosing the Threshold for Large Covariance Estimations in Factor Models}
\author{Yuan Liao\footnote{Department of Economics,  Rutgers University, New Brunswick, NJ 08901.}\medskip\\{Rutgers University}   }
\maketitle
 \begin{abstract}
 This note shows that for i.i.d. data, estimating large covariance matrices in factor models can be casted using a simple plug-in method to choose the threshold:
$$ \mu_{jl}=\frac{c_0}{\sqrt{n}}\Phi^{-1}(1-\frac{\alpha}{2p^2})\sqrt{\frac{1}{n}\sum_{i=1}^n\widehat u_{ji}^2\widehat u_{li}^2}.$$
This is motivated by the tuning parameter suggested by \cite{belloni2012sparse} in the lasso literature.
  It also leads to the minimax rate of convergence of the large covariance matrix estimator.  Previously, the minimaxity is achievable only when $n=o(p\log p)$ by \cite{POET}, and now this condition is weakened to  $n=o(p^2\log p)$. Here $n$ denotes the sample size and $p$ denotes the dimension.

 \end{abstract}

\onehalfspacing

\section{Introduction}
 
 Estimating large covariance matrices in factor models has been an important research area in recent years. This note provides a practical guidance on how to choose the threshold value for the method introduced in  \cite{POET}. Consider a simple factor model:
 $$
y_{ji}=\lambda_j'f_i+u_{ji},\quad i\leq n, j\leq p,
 $$
 where $\lambda_j$ denotes a $K$-dimensional vector of loadings for the $j$ th individual, and $f_i$ is a vector of common factors for the $i$ th observation.  In this model, only $y_{ji}$ is observable.  In the usual notation for factor models, the dependent variable is often denoted by $y_{it}$ for the $i$ the individual observed at time $t$.  In this note, however, we shall apply the moderate deviation theory for self-normalized sequences of independent data \citep{de2009self}. Therefore, we stick to the more traditional notation in the statistical literature, and use $j\leq p$ to denote the index of variables and $i\leq n$ as the observations. That being said, the serial correlation is ruled out, due to the technical tools we are using to approximate the distribution of self-normalized sums.
 
 The object of interest is to estimate the $p\times p$ covariance matrix of $u_n=(u_{1n},...,u_{pn})'$, denoted by $\Sigma_u$, from the observations $\{y_{ji}\}_{j\leq p, i\leq n}$. Estimating $\Sigma_u$  leads to  many interesting applications. First, it makes it possible to obtain a good estimate of the covariance for $y_i=(y_{1i},...,y_{pi})'$. Secondly, it allows as to improve the estimation of factors and loadings in the presence of  cross-sectional correlations \citep{choi, BL13}. Third, it also ``activates" many classical Wald statistics for high-dimensional testing problems, which otherwise cannot handle the difficulty of using a large inverse weight matrix \citep{fan2015power}.
  In the so-called ``approximate factor models", this is a large and non-diagonal covariance. 
 
 I now  describe \cite{POET}'s estimator. Let $\widehat f_i$ and $\widehat \lambda_j$ respectively denote the factors and loading estimators, which can be obtained via, e.g., the principal components method. Then we obtain the residual estimate $\widehat u_{ji}=y_{ji}-\widehat\lambda_j'\widehat f_i$. The residual sample covariance is then
 $$
 \widehat S_u=(\widehat s_{jl})_{p\times p},\quad \widehat s_{jl}=\frac{1}{n}\sum_{i=1}^n\widehat u_{ji}\widehat u_{li}.
 $$
 Next, we apply soft-thresholding to obtain $\widehat\Sigma_u=(\widehat\sigma_{jl})_{p\times p}$, where
 $$
 \widehat\sigma_{jl}=\begin{cases}
\widehat s_{jl},& j=l\\
 sgn(\widehat s_{jl})(|\widehat s_{jl}|-\mu_{jl})_+,& j\neq l.
 \end{cases}
 $$
 Here sgn($\widehat s_{jl}$) denotes the sign of $\widehat s_{jl}$, and $(x)_+=\max(x, 0)$. What plays the central role is the user-specified thresholding value $\mu_{jl}$, which may depend on $(n,p)$, but we suppress such dependence in the notation. More importantly, its dependence on $(j, l)$ indicates that the threshold should not be chosen as a universe constant. Ideally, it should be the smallest constant that just dominates the statistical error
 $
 |\widehat s_{jl}-Eu_{ji}u_{li}|.
 $
 \cite{POET} took the correlation matrix as their standpoint, which left a constant unspecified in the thresholding value, and suggested choose it by the cross-validation. In addition, the choice of the constant also depends on the number of factors to use in the model. \footnote{I communicated with Michael Wolf on this procedure when applied to the portfolio selection problems. Michael implemented it on  the daily returns of 252 trading days, with the number of stocks varying from 30 through 500, and  suggested using the constant ``one" with five factors.}
 
 This note suggests, in contrast, a simple plug-in choice for $\mu_{jl}$ as follows:
 \begin{equation}\label{e1}
\mu_{jl}=\frac{c_0}{\sqrt{n}}\Phi^{-1}(1-\frac{\alpha}{2p^2})\sqrt{\frac{1}{n}\sum_{i=1}^n\widehat u_{ji}^2\widehat u_{li}^2}.
 \end{equation}
 Here $c_0>1$ is taken as a constant arbitrarily close to one, e.g., $c_0=1.1$. 
This is equivalent to thresholding the studentized $\widehat s_{jl}$ using a universe constant $c_0\Phi^{-1}(1-\frac{\alpha}{2p^2})$. Here $\Phi^{-1}$ denotes the inverse standard normal CDF; $\alpha$ is a small significant level so that 
$$
P(\max_{jl\leq p} |\widehat s_{jl}-Eu_{ji}u_{li}|/\mu_{jl}>1)\leq \alpha+o(1).
$$
We can choose, e.g.,  $\alpha=0.05$.  Everything else in the definition of $\mu_{jl}$ is completely data-driven, and is easy to plug in. This method uses the fact that the  distribution of the normalized average
$$
|\frac{1}{n}\sum_{i=1}^nu_{ji}u_{li}|/ \sqrt{\frac{1}{n}\sum_{i=1}^nu_{ji}^2u_{li}^2}
$$
can be well approximated by the standard normal distribution, and was previously studied  by  \cite{Cai11b} for estimating sparse covariances.  The idea was also used commonly in the  high-dimensional lasso literature (\cite{belloni2012sparse,  belloni2014inference}).

It is helpful to look at the components of $\widehat s_{jl}$ more carefully. In fact,
$$
\widehat s_{jl}-Eu_{ji}u_{li}= \underbrace{\frac{1}{n}\sum_{i=1}^n\widehat u_{ji}\widehat u_{li}-\frac{1}{n}\sum_{i=1}^nu_{ji}u_{li}}_{\text{estimating residuals}} + \underbrace{\frac{1}{n}\sum_{i=1}^nu_{ji}u_{li}-Eu_{ji}u_{li}}_{\text{empirical process}}.
$$
The plug-in thresholding value in fact bounds the empirical process components uniformly over all the individuals. But it does not control the ``estimating residuals" components. Using the Cauchy-Schwarz inequality, \cite{POET} showed that this component has a rate of convergence $O_P(\frac{1}{\sqrt{p}}+\sqrt{\frac{\log p}{n}})$, hence cannot be ignored. In fact, we will see  that this rate can be improved to 
$$
\max_{jl\leq p}|\frac{1}{n}\sum_{i=1}^n\widehat u_{ji}\widehat u_{li}-\frac{1}{n}\sum_{i=1}^nu_{ji}u_{li}|=O_P(\frac{\log p}{n}+ \frac{1}{p}).
$$
Therefore, it is negligible so long as $n=o(p^2\log p)$.  

On the other hand, when $n=o(p^2\log p)$ is not satisfied,   the plug-in choice would under-threshold the residual sample covariance. The amount of under-thresholding would be at most
$
O_P(\frac{1}{p^2})\times p^2
$
when the rate of convergence is with respect to the  squared frobenius norm.

In this note, we write $\|A\|_{\max}=\max_{ij}|A_{ij}|$, and $\|A\|=\sqrt{\nu_{\max}(A'A)}$, where $\nu_{\max}(A)$ denotes the maximum eigenvalue of $A$. We state all the propositions  without providing their proofs. The technical proofs follow from standard arguments in this literature, and are available upon requests. 

\section{Identification}

 As we only observe $y_{ji}$, consistently estimating $\Sigma_u$ in any reasonable sense is  possible only  if $u_i$ can be approximately identified from the factor components.  Write $Y$  and $U$  the $p\times n$ matrices of $y_{ji}$ and $u_{ji}$. Write $\Lambda$ as the $p\times K$ matrix of $\lambda_j$ and $F$  as the $n\times K$ matrix of $f_i$. Then the matrix form of the model is 
 $$
 Y=\Lambda F'+U.
 $$
 This implies 
 $
 YY'=\Lambda F'F\Lambda'+\Lambda F'U'+(\Lambda F'U')'+UU'.
 $
Now take the expectation on both sides yields 
$$
\frac{1}{pn}E(YY')=\frac{1}{p}\Lambda Ef_if_i'\Lambda'+\frac{1}{pn}E(UU').
$$
 We now see that the ``pervasive condition" plays a central  role in the identification: 
 $$
\text{The minimum eigenvalue of } \Lambda'\Lambda \text{ dominates the maximum eigenvalue of  $\Sigma_u$}.
 $$
Further suppose all the eigenvalues of $Ef_if_i'$ are bounded away from zero. Given these conditions,  as $p\to\infty$, the first $K$ eigenvalues of $\frac{1}{p}\Lambda Ef_if_i'\Lambda'$ do not vanish, but all the eigenvalues of $\frac{1}{pn}E(UU')$ decays to zero.
 
 In addition, right multiplying $\Lambda$ yields 
 $$
[ \frac{1}{pn}E(YY')-\frac{1}{pn}E(UU')]\Lambda=\Lambda Ef_if_i' \frac{1}{p}\Lambda'\Lambda
 $$
 Hence there is a $K\times K$ matrix $H$ so that the  columns of $\Lambda H$ are the first $K$ eigenvectors of $\frac{1}{pn}E(YY')-\frac{1}{pn}E(UU')$, which are then approximately the first $K$ eigenvectors of $\frac{1}{pn}E(YY')$ since $\frac{1}{pn}E(UU')$ is dominated as $p\to\infty$, due to the sin-theta theorem. 
  Hence, as the dimension diverges, the pervasive condition ensures that $\Lambda$ can be identified up to a rotation matrix. 
  
  Now left multiplying $ \frac{1}{p}H'\Lambda'$   on both sides of $Y=\Lambda F'+U$   yields:
 $$
 H^{-1}F'=(\frac{1}{p}H'\Lambda'\Lambda H)^{-1} \frac{1}{p}H'\Lambda'Y-(\frac{1}{p}H'\Lambda'\Lambda H)^{-1} H' \frac{1}{p}\Lambda'U.
 $$
 The second term on the right hand side is dominated under suitable conditions that $\frac{1}{p}\Lambda'U$ is ``smaller" than  $\frac{1}{p}\Lambda' \Lambda F'$.  Therefore, the asymptotic identification of $\Lambda H$ yields the asymptotic identification of $H^{-1}F'$, as well as that of 
 $$
 U=Y-\Lambda HH^{-1}F'.
 $$
 
 The following theorem gives a  formal identification result using the $\|.\|_{\max}$ norm.
 \begin{prop}\label{p2.1}
Let $\nu_1\geq...\geq\nu_p$ be the eigenvalues of $\frac{1}{n}E(YY')$, and let    $\xi_1,...,\xi_p$ be the corresponding eigenvectors. Suppose  all the eigenvalues of $\Lambda'\Lambda/p$  are bounded away from zero and infinity, and $\|\Sigma_u\|=O(1)$. Then
$$
\|  \Sigma_u- \sum_{l=K+1}^p\nu_l\xi_l\xi_l'\|_{\max}=O(\frac{1}{\sqrt{p}}).
$$
 \end{prop}
 We omit the proof of this proposition in this note. Heuristically, the key step is to prove $\|\frac{1}{p}\Lambda Ef_if_i'\Lambda'-\sum_{l=1}^K\nu_l\xi_l\xi_l'\|_{\max}=O(\frac{1}{\sqrt{p}})$, by showing $\|\xi_l-(\Lambda H)_l \|(\Lambda H)_l\|^{-1}  \|=O(p^{-1})$ and $|\nu_l- \|(\Lambda H)_l\|^2 |=O(\|\Sigma_u\|)$ for some $K\times K$ matrix $H$ and $l\leq K$, where $(\Lambda H)_l$ denotes the $l$ th column of $\Lambda H$. 
   Then this proposition follows from the decompositions
 $
\frac{1}{n}E(YY')=\Lambda Ef_if_i'\Lambda'+\Sigma_u,
$
and $\frac{1}{n}E(YY')= \sum_{l=1}^K\nu_l\xi_l\xi_l'+ \sum_{l=K+1}^p\nu_l\xi_l\xi_l'$.

\section{Improved Rate of Convergence}
We employ the PC estimator of \cite{bai03, POET} to estimate $F,\Lambda$ and $U$. 
Recall that $$
\widehat s_{jl}-Eu_{ji}u_{li}= \underbrace{\frac{1}{n}\sum_{i=1}^n\widehat u_{ji}\widehat u_{li}-\frac{1}{n}\sum_{i=1}^nu_{ji}u_{li}}_{\text{estimating residuals}} + \underbrace{\frac{1}{n}\sum_{i=1}^nu_{ji}u_{li}-Eu_{ji}u_{li}}_{\text{empirical process}}.
$$
\subsection{Empirical process}

Using the moderate deviation theory for self-normalized sequences \citep{de2009self}, the second difference on the right hand side is bounded by 
$$
 \frac{1}{\sqrt{n}}\Phi^{-1}(1-\frac{\alpha}{2p^2})\sqrt{\frac{1}{n}\sum_{i=1}^nu_{ji}^2u_{li}^2}\leq \mu_{jl}\frac{c_0+1}{2c_0}
$$
with probability at least $1-\alpha-o(1)$, uniformly in $j,l\leq p$. Here  $c_0>1$ and $\mu_{jl}$ are  defined in Introduction.  Besides that the data are  independent, the sufficient conditions are,  for some $c,C>0$,
\begin{equation}\label{e3.1}
\min_{jl\leq p}Eu_{ji}^2u_{li}^2>c,\quad \max_{jl\leq p}Eu_{ji}^6<C,
\end{equation}
and
\begin{equation}\label{e3.2}
\max_{jl}|\frac{1}{n}\sum_{i=1}^n(u_{ji}^2u_{li}^2-\widehat u_{ji}^2\widehat u_{li}^2)|=o_P(1).
\end{equation}
In addition, it is required that $\log p=o(n^{1/3}).$

\subsection{Estimating residuals}

 We provide additional regularity conditions.
 \begin{assum}\label{a3.1}
(i) $\{f_i, u_i\}_{i\leq n}$ are i.i.d. and sub-Gaussian.\\
(ii) $E(u_i|f_i)=0$.\\
(iii) There is $C>0$, $\max_{j\leq p}\|\lambda_j\|<C$.\\ 
(iv) All the eigenvalues of $\frac{1}{p}\Lambda'\Lambda$ are bounded away from both zero and infinity.
 \end{assum}
 
 The following assumption requires the weak cross-sectional correlations.
  \begin{assum}\label{a3.2} There is $C>0$,\\
(i) $\max_{l\leq p}\frac{1}{np}\sum_{i=1}^n\sum_{j,m\leq p}|\cov(u_{mi}u_{li}, u_{ji}u_{li})|<C.
$ \\
(ii) $\max_{l\leq p}\sum_{j=1}^p|Eu_{ji}u_{li}|<C$. \\
(iii)  $E(\frac{1}{\sqrt{p}}\sum_{j=1}^pu_{ji}u_{jk}-Eu_{ji}u_{jk})^4<C$.\\
(iv) $E\|\frac{1}{\sqrt{p}}\sum_{j=1}^p\lambda_{j}u_{ji}\|^4<C$.

 \end{assum}

\begin{prop}\label{p3.1}
Under Assumptions \ref{a3.1}, \ref{a3.2} and those of Proposition \ref{p2.1}, we have
$$
\max_{jl\leq p}|\frac{1}{n}\sum_{i=1}^n\widehat u_{ji}\widehat u_{li}-\frac{1}{n}\sum_{i=1}^nu_{ji}u_{li}|
=  O_P(\frac{1}{{p}}+{\frac{\log p}{n}}).
$$
\end{prop}

We omit the proof of this proposition. Heuristically speaking, it can be shown by directly computing the expansions of the PC estimators  $\widehat\lambda_{j}$ and $\widehat f_i$,
$$
\max_{jl\leq p}|\frac{1}{n}\sum_{i=1}^n\widehat u_{ji}\widehat u_{li}-\frac{1}{n}\sum_{i=1}^nu_{ji}u_{li}|=O_P(\frac{\log p}{n})+\max_{j\leq p}|\frac{1}{n}\sum_{i=1}^n(H^{'-1}f_i-\widehat f_i)u_{ji}|
$$
for some rotation matrix $H$. Importantly, \cite{POET} used the Cauchy-Schwarz inequality to bound the second term on the right hand side, and reached a non-sharp rate of convergence:
$
O_P(\frac{1}{\sqrt{p}}+\frac{1}{\sqrt{n}}).
$
In fact, proposition \ref{p3.1} can be proved by showing that 
$$
\max_{j\leq p}|\frac{1}{n}\sum_{i=1}^n(H^{'-1}f_i-\widehat f_i)u_{ji}|=O_P(\frac{1}{n}+\frac{1}{p}+\sqrt{\frac{\log p}{np}}).
$$

 For some $q\in[0,1)$, define
 $$
 m_p=\max_{j\leq p}\sum_{l=1}^p|Eu_{ji}u_{li}|^q.
 $$

Proposition \ref{p3.1} leads to two exciting results.  One is an improved rate of convergence of the covariance matrix estimation. Note that 
the convergence rate of  $\|\widehat\Sigma_u-\Sigma_u\|$ is determined by    that of $\max_{jl\leq p}|\frac{1}{n}\sum_{i=1}^n\widehat u_{ji}\widehat u_{li}-Eu_{ji}u_{li}|$.   \cite{POET} showed that (in the special case $q=0, m_p=O(1)$), $\|\widehat\Sigma_u-\Sigma_u\|=O_P(\frac{1}{\sqrt{p}}+\sqrt{\frac{\log p}{n}})$. Due to Proposition \ref{p3.1}, this rate can be sharpened to
$$\|\widehat\Sigma_u-\Sigma_u\|=O_P(\frac{1}{{p}}+\sqrt{\frac{\log p}{n}}
).$$
The impact of estimating the unknown factors, is therefore weakened to  $O_P(\frac{1}{p})$. The other exciting implication is that it is now possible to use the simple plug-in choice $\mu_{jl}$ as the thresholding value, so long as $n=o(p^2\log p)$.  In this case $\max_{jl\leq p}|\frac{1}{n}\sum_{i=1}^n\widehat u_{ji}\widehat u_{li}-\frac{1}{n}\sum_{i=1}^nu_{ji}u_{li}|=o_P(\sqrt{\frac{\log p}{n}})$, hence is a smaller order than the empirical process part.   Consequently, with probability at least $1-\alpha-o(1)$,
$$
|\widehat s_{jl}-Eu_{ji}u_{li}|\leq \mu_{jl}
$$
uniformly in $j,l\leq p$.  Hence for  the sparse covariance $\Sigma_u$, most of the elements of the estimated sample residual covariance are dominated by $\mu_{jl}$.

\begin{cor} Use the thresholding value $\mu_{jl}$ given by (\ref{e1}).
When $n=o(p^2\log p)$ and $\log p=o(n^{1/3})$, under the assumptions of  Proposition \ref{p3.1}, and conditions (\ref{e3.1}) (\ref{e3.2}),
$$\|\widehat\Sigma_u-\Sigma_u\|=O_P( m_p(\frac{\log p}{n})^{(1-q)/2}
).$$
\end{cor}
This is the minimax convergence rate. 

\section{Conclusion}

In this note, I sharpen the rate of convergence of $\max_{jl\leq p}|\frac{1}{n}\sum_i\widehat u_{ji}\widehat u_{li}-Eu_{ji}u_{li}|$ for the PC estimator in approximate factor models. The rate is faster than that of \cite{POET} when $n=o(p^2\log p)$.  An immediate consequence  is a simple plug-in choice of the thresholding value for the estimated idiosyncratic covariance matrix, which takes a similar type to that of   \cite{belloni2012sparse} in the lasso literature. It also leads to the minimax rate of convergence of the large covariance matrix estimator.  Previously, the minimaxity is achievable only when $n=o(p\log p)$ by \cite{POET}, and now this condition is weakened to  $n=o(p^2\log p)$.

 \newpage
  \bibliographystyle{ims}
\bibliography{liaoBib}

\end{document}